\documentstyle[twocolumn,aps,prl]{revtex}
\newcommand{\g}[1]{{\bf {#1}}}
\begin{document}

\title{Slow flows of an relativistic perfect fluid 
in a static gravitational field}
\author{V.P. Ruban \cite{1}}
\address {\it Optics and Fluid Dynamics Department, 
Ris\o ~National Laboratory, DK-4000 Roskilde, Denmark \\
and  \\
L.D.Landau Institute for Theoretical Physics,
2 Kosygin str., 117334 Moscow, Russia}
\maketitle

\begin{abstract}
Relativistic hydrodynamics of an isentropic fluid in a gravitational field
is considered as the particular example from the family of Lagrangian
hydrodynamic-type systems which possess an infinite set of integrals of
motion due to the symmetry of Lagrangian with respect to relabeling of
fluid particle labels. Flows with fixed topology of the vorticity are
investigated in quasi-static regime, when deviations of the space-time
metric and the density of fluid from the corresponding equilibrium
configuration are negligibly small. On the base of the variational 
principle for frozen-in vortex lines dynamics, the equation of motion for 
a thin relativistic vortex filament is derived in the local induction
approximation.
\end{abstract}

\medskip


\medskip


{\bf 1.} In the given work inviscid flows of a relativistic isentropic
fluid \cite{LL6} are investigated in frame of the general relativity theory
\cite{LL2}, i.e. in significantly curved space-time parameterized by some
coordinate system $(t,\g{r})$, with the metric tensor $g_{ik}(t,\g{r})$.

For any hydrodynamic system it is possible to separate vortex and sound
degrees of freedom, with the vortices being characterized as the "soft", 
while the sound as the "hard" degrees of freedom. If there is a static
equilibrium state in the system with the density of fluid 
$\rho=\rho_0(\g{r})$ and with the velocity $\g{v}=0$, then the dynamical
regime is possible with the hard degrees of freedom being excited weakly
and with dynamics being entirely described by slow flows on the background
of practically unchanged density  $\rho_0(\g{r})$. A typical velocity in
such flows should be much smaller then the sound speed (and also smaller
then the speed of light, as in the case). The general theory of such slow
vortical flows in spatially inhomogeneous systems has been developed
recently by the present author in his work \cite{Ruban2000}. In particular,
the Hamiltonian equations of motion have been obtained for an arbitrary
given topology of the canonical vorticity field, and the general form of
the Lagrangian for frozen-in vortex lines has been established. 

The goal of this Brief Report is to apply the formalism developed to
the model of relativistic hydrodynamics. As the equilibrium state, a
spherically symmetric static distribution of a fluid will be considered
which is the reason for space-time curvature. In the expression for the
action of relativistic hydrodynamics \cite{Brown} the only arbitrary
dependence $\varepsilon(n)$ appears which connects the density
$\varepsilon$ of the total energy of fluid measured in the locally
co-moving reference frame with the density $n$ of number of conserved
particles \cite{LL6}. The scalar $n$ is equal to the absolute value of the
current 4-vector $n^i=n(dx^i/ds) \cite{LL2}$. Therefore the static metric
of the space-time, as well as the equilibrium distribution $\rho_0(\g{r})$
of the matter in space, after the equation of state $\varepsilon(n)$ has
been fixed, depend on single parameter -- the total amount of fluid in the
system.

As the result, the equation of motion of a slender vortex filament in
relativistic fluid will be derived in the local induction approximation,
and it will become clear how the the dynamics of frozen-in vorticity in
curved space-time differs from the non-relativistic hydrodynamics, even in
the case of small velocities.

{\bf 2.} Before starting the consideration of the concrete dynamical
system, it is necessary to make some preliminary remarks. The ideal
relativistic hydrodynamics is a member of the large family of the
hydrodynamic-type Lagrangian models. This family is remarkable because of
many interesting properties, first of all due to presence of infinite
number of specific integrals of motion. Therefore it is reasonable in the
beginning to consider the relativistic hydrodynamics from general positions
of the canonical formalism \cite{Arnold},\cite{DNF}.

As known, the entire Lagrangian description of the flow of some continuous
medium can be given by the 3D mapping $\g{r}=\g{x}(t,\g{a})$, which
indicates the space coordinates of each medium point labeled by a label
$\g{a}=(a_1,a_2,a_3)$, at an arbitrary moment in time $t$. The labeling
$\g{a}$ can be chosen in such a manner that the amount of matter in a
small volume $d^3\g{a}$ in the label space is simply equal to this volume.
Inasmuch as the dynamical system is supposed to be conservative, the
equations of motion for the mapping $\g{x}(t,\g{a})$ (as well as for the
gravitational field, as in our case) follow from a variational principle
\begin{equation}\label{min_action}
\delta S=\delta\int{\cal L}\{\g{x}(t,\g{a}),g_{ik}(t,\g{r})\}dt=0,
\end{equation}
where the Lagrangian ${\cal L}$ is a functional of $\g{x}(t,\g{a})$,
$g_{ik}(t,\g{r})$ and their derivatives. The explicit expression for 
${\cal L}$ will be given later, but now let us note the very important
circumstance related to the fluidity property of the medium under
consideration. The fluidity is manifested in the fact that the Lagrangian
actually contains the dependence on $\g{x}(t,\g{a})$ only through two
Eulerian characteristics of the flow, namely through the field of relative
density $\rho(t,\g{r})$ and the velocity field $\g{v}(t,\g{r})$, i.e.
${\cal L}={\cal L}\{\rho,\g{v},g_{ik}\}$, with
\begin{equation}\label{rho_def}
\rho(t,\g{r})=\frac{1}{\mbox{det}\|\partial\g{x}/\partial\g{a}\|}
\Big|_{\g{a}=\g{x}^{-1}(t,\g{r})}\quad,
\end{equation}
\begin{equation}\label{v_def}
\g{v}(t,\g{r})=\dot{\g{x}}(t,\g{a})|_{\g{a}=\g{x}^{-1}(t,\g{r})}
\end{equation}
It follows from these definitions that the dynamics of the "density"
$\rho(t,\g{r})$ obeys the continuity equation in its standard
form
\begin{equation} \label{rho_t}
\rho_t+\nabla(\rho{\bf v})=0.
\end{equation}

The vanishing condition for variation of the action
$S=\int{\cal L}\{\rho,\g{v},g_{ik}\}dt$ when the mapping $\g{x}(t,\g{a})$
is varied by $\delta\g{x}(t,\g{a})$, with taking into account the obvious
relations between the variations
\begin{equation}\label{rho_var}
\delta\rho(\g{r})=-\nabla(\rho(\g{r})\cdot\delta\g{x}(\g{a}(\g{r}))),
\end{equation}
\begin{equation}\label{v_var}
\delta\g{v}(\g{r})=\delta\dot{\g{x}}(\g{a}(\g{r}))-
(\delta\g{x}(\g{a}(\g{r}))\cdot\nabla)\g{v},
\end{equation}
can be expressed in Eulerian representation as follows (the generalized
Euler equation \cite{R99})
\begin{equation} \label{dynequation} 
(\partial_t+{\bf v\cdot\nabla})
\left(\frac{1}{\rho}\cdot\frac{\delta {\cal L}}{\delta {\bf v}}\right)=
\nabla\left(\frac{\delta {\cal L}}{\delta \rho}\right)-
\frac{1}{\rho}\left(\frac{\delta {\cal L}}{\delta v^\alpha}\right)
\nabla v^\alpha.
\end{equation}
Together with the conditions $\delta S/\delta g_{ik}=0$, the equations
(\ref{rho_t}) and (\ref{dynequation}) determine completely the evolution of
hydrodynamic system. It is very important that in all such systems an
infinite number of conservation laws exists \cite{Salmon}-\cite{IL}. 
The reason of their existence is that the Lagrangian 
${\cal L}\{\g{x}(t,\g{a}),g_{ik}(t,\g{r})\}$ admits the infinite-parametric
symmetry group -- it assumes the same value on any two mappings 
$\g{x}_1(t,\g{a})$ and $\g{x}_2(t,\g{a})$, if they differ one from another
only by some relabeling of the labels 
\begin{equation} \label{relabeling}
\g{x}_2(t,\g{a})=\g{x}_1(t,\g{a}^*(\g{a})), \qquad
\mbox{det}\|\partial\g{a}^*/\g{a}\|=1.
\end{equation}
Obviously, such mappings create the same "density" and velocity fields.
According to the Noether's theorem \cite{Arnold},\cite{DNF},  every
one-parametric sub-group of the relabeling group $\g{a}^*(\g{a})$ with
unit Jacobian corresponds to an integral of motion. There are several
classifications of these conservation laws. For instance, one can postulate
that the circulation of the canonical momentum $\g{p}(t,\g{r})$
\begin{equation} \label{p_def}
\g{p}(t,\g{r})\equiv\frac{\delta{\cal L}}{\delta\dot\g{x}(\g{a}(t,\g{r}))}
=\frac{1}{\rho}\left(\frac{\delta {\cal L}\{\rho,\g{v},g_{ik}\}}
{\delta {\bf v}}\right)
\end{equation}
along an arbitrary frozen-in closed contour $\gamma(t)$ does not depend on
time (the generalized theorem of Kelvin)
$$
\oint_{\gamma(t)}(\g{p}\cdot d\g{r})=const.
$$
We arrive at a different formulation when consider the solenoidal field of
the vorticity $\g\Omega(t,\g{r})$
\begin{equation} \label{Omega_def}
\g{\Omega}(t,\g{r})=\mbox{curl}\,\g{p}(t,\g{r}).
\end{equation}
From the equation (\ref{dynequation}) the equation of frozenness for
$\g\Omega(t,\g{r})$ follows
\begin{equation} \label{Omega_motion}
\g{\Omega}_t=\mbox{curl}[\g{v}\times\g{\Omega}].
\end{equation} 
The formal solution of this equation is
$$
{\bf\Omega}(t,{\bf r})=
\frac{{\bf (\Omega_0({\bf a})\nabla_a)x}(t,\g{a})}
{\mbox{det}\|\partial{\bf x}(t,{\bf a})/\partial{\bf a}\|}
\Big|_{{\bf a}={\bf x}^{-1}(t,{\bf r})}=
$$
\begin{equation} \label{Om_r}
=\int\delta({\bf r}-{\bf x}(t,{\bf a}))
({\bf\Omega}_0({\bf a})\nabla_{\bf a}){\bf x}(t,{\bf a})d{\bf a},
\end{equation}
with the solenoidal field ${\bf\Omega}_0({\bf a})$ independent on time
being exactly an integral of motion (the Cauchy invariant \cite{Lamb}).
The formula (\ref{Om_r}) displays that the lines of the initial field 
${\bf \Omega_0}$ are transported by the flow, retaining all the topological
characteristics. Such a property of vortex lines is known as their
frozenness.

{\bf 3.} Let's turn our attention to the relativistic hydrodynamics and
consider the corresponding expression for the action $S$ 
\cite{Brown}, \cite{LL2}
\begin{equation} \label{S_rel}
S=-\int\sqrt{-g}\left(\varepsilon(n)+\frac{1}{16\pi}
R[g_{ik}]\right)d\g{r}dt.
\end{equation}
Here $g=\mbox{det}\|g_{ik}\|$ is the determinant of the metric tensor,
$R[g_{ik}]$ is the scalar curvature of space-time \cite{LL2}. In this
formula, however, the scalar $n$ should be expressed through the dynamical
variables $\{\rho,\g{v},g_{ik}\}$. It is easy to see, comparing the
continuity equation (\ref{rho_t}) with the equation of fluid conservation
\cite{LL6}
\begin{equation}
n^i_{;i}\equiv\frac{1}{\sqrt{-g}}\frac{\partial}{\partial x^i}
\left(\sqrt{-g}n\frac{dx^i}{ds}\right)=0,
\end{equation}
that $\rho=\sqrt{-g}n({dt}/{ds})$. From this we have the relation
\begin{equation} \label{n_rho_g_v}
n=\frac{\rho}{\sqrt{-g}}
\sqrt{g_{00}+g_{0\alpha}v^\alpha+g_{\alpha\beta}v^\alpha v^\beta},
\end{equation}
which should be substituted into the expression (\ref{S_rel}) and after 
that the equations of motion can be obtained in accordance with equations
(\ref{rho_t}), (\ref{dynequation}) and $\delta S/\delta g_{ik}=0$. In such
form, however, the action contains an arbitrariness related to the
possibility of coordinate changing. Therefore four gauge conditions can be
imposed on the metric tensor \cite{LL2}. Even more, inasmuch as we are
interested in a static central-symmetric gravitational field, the
corresponding metric can be chosen as follows
\begin{equation} \label{stat_metric}
(ds^2)^{stat}=A(r)dt^2-B(r)(dx^2+dy^2+dz^2),
\end{equation}
where $r=\sqrt{x^2+y^2+z^2}$. The expression for the scalar curvature in
this metric takes the form
$$
R^{stat}[A,B]=\frac{1}{2 A^2 B^3}\Big(
-B^2A'^2+
$$
$$
+AB(A'B'+2B(2A'/r+A''))+
$$
\begin{equation} \label{curvature}
+A^2(-3B'^2+4B(2B'/r+B''))
\Big).
\end{equation}
Therefore the static state is determined by an extremum of the functional
\begin{equation} \label{S_stat}
\frac{S^{stat}}{4\pi}=-\int\sqrt{AB^3}\left[
\varepsilon\left(\frac{\rho_0}{\sqrt{B^3}}\right)+
\frac{R^{stat}[A,B]}{16\pi}\right]r^2dr dt.
\end{equation}
The variations on $\delta A(r)$ and on $\delta B(r)$ should be performed in
the standard way when an extremum is being found, while the variation on
displacements $\delta\g{x}(\g{a})$, with taking into account the relation
(\ref{rho_var}) and subsequent integration, gives the relativistic 
hydrostatic equation for isentropic fluid \cite{LL6}
\begin{equation} \label{hydrostat}
\sqrt{A}w\left(\frac{\rho_0}{\sqrt{B^3}}\right)=\lambda=const,
\end{equation}
where $w(n)=\varepsilon'(n)$ is the relativistic enthalpy per one
conservative particle, while a constant of integration $\lambda$ is related
to the total amount of matter in the system.

With fixed static central-symmetric metric (\ref{stat_metric}) the
Lagrangian of relativistic hydrodynamics takes the form
\begin{equation} \label{L_r}
{\cal L}_r\{\rho,\g{v}\}=-\int\sqrt{AB^3}\,\,
\varepsilon\left(\frac{\rho\sqrt{A-B\g{v}^2}}{\sqrt{AB^3}}\right)
d\g{r}.
\end{equation}

As far as we examine here only slow flows, we need only the first term from
the expansion of ${\cal L}_r$ on powers of $v^2$
$$
{\cal L}_*=\int w\left(\frac{\rho_0}{\sqrt{B^3}}\right)
\cdot\frac{\rho_0 B}{\sqrt{A}}\cdot\frac{\g{v}^2}{2}d\g{r}=
\lambda \int \frac{\rho_0 B}{A}\cdot\frac{\g{v}^2}{2}d\g{r}
$$
Just the functional ${\cal L}_*$ determines the slow dynamics of frozen-in
vorticity $\g{\Omega}$
$$
\g{\Omega}\approx\mbox{curl}\left(
\frac {1}{\rho_0}\frac{\delta{\cal L}_*}{\delta \g{v}}\right)=
\mbox{curl}\left(\lambda\frac{B}{A}\g{v}\right).
$$

{\bf 4.} In order to investigate the motion of frozen-in vortex lines in
the given problem, it is necessary to define the Hamiltonian ${\cal H}_*$
\cite{Ruban2000}
\begin{equation}\label{Hdef}
{\cal H}_*=
\int\Big(\frac{\delta{\cal L}_*}{\delta {\bf v}}
\cdot{\bf v}\Big)d{\bf r}-{\cal L}_*=
\frac{1}{\lambda}\int\frac{\rho_0 A}{B}\cdot\frac{\g{p}^2}{2}d\g{r}
\end{equation}
and then express it through the vorticity $\g{\Omega}$ with help of
the relation (\ref{Omega_def}) and the condition of zero variation of 
density $\rho_0$
\begin{equation}\label{density}
\nabla\left(\frac{\delta {\cal H}_*}{\delta {\bf p}}\right)=
\nabla\left(\frac{A\rho_0}{\lambda B}\g{p}\right)=0.
\end{equation}
The equation of motion for $\g{\Omega}$ in quasi-static regime can be
written after that in the form \cite{Ruban2000}
\begin{equation}\label{Ham}
{\bf\Omega}_t=\mbox{curl}
\left[\mbox{curl}\left(\frac{\delta{\cal H}_*}{\delta{\bf\Omega}}\right)
\times\frac{{\bf\Omega}}{\rho_0(\g{r})}
\right].
\end{equation}
Inasmuch as the vortex lines are frozen into the fluid, it is reasonable to
consider their shape as the new dynamical object \cite{Ruban2000} and
parameterize the vorticity field as follows
\begin{equation}\label{lines}
{\bf \Omega }({\bf r},t)=\int_{\cal N}d^2\nu \oint \delta ({\bf r}-
{\bf R}(\nu,\xi,t)){\bf R}_{\xi}d\xi.  
\end{equation}
Here the label $\nu=(\nu_1,\nu_2)\in{\cal N}$ belongs to the 2D manifold
${\cal N}$ and singles out a vortex line, while the parameter $\xi$
determines a point on the line. Obviously, the choice of the longitudinal
parameter $\xi$ is not unique \cite{KR2000}. It is easy to understand the
meaning of the above formula -- the frozen-in vorticity field is
represented here as the continuous distribution of vortex lines.

It follows from the equations (\ref{Ham}) and (\ref{lines}) that the 
equation of motion of vortex lines has the Hamiltonian form
(see \cite{Ruban2000} for detailed derivation)
\begin{equation}\label{main}
\left[ {\bf R}_\xi\times
{\bf R}_{t}\right]\rho_0(\g{R}) =
\frac{\delta {\cal H}_*\{{\bf \Omega }\{{\bf R}\}\}}
{\delta {\bf R}}.  
\end{equation}
Clearly, this equation does not depend on the choice of longitudinal
parameterization. It is easy to check by direct calculation that the above
equation corresponds to the variational principle 
$\delta\int{\cal L}_{\cal N}dt=0$ with the Lagrangian
${\cal L}_{\cal N}$ of the form \cite{Ruban2000}
$$
{\cal L}_{\cal N}=
\int_{\cal N}d^2\nu\oint \Big(\left[ {\bf R}_{t}(\nu,\xi)\times
{\bf D}({\bf R}(\nu,\xi))\right]
\cdot
{\bf R}_\xi(\nu,\xi)\Big)d\xi 
$$
\begin{equation}\label{L_N}
 - {\cal H}_*\{{\bf \Omega }\{{\bf R}\}\},  \label{LAGR_lines}
\end{equation}
Here the vector function ${\bf D}({\bf R})$ is related to the equilibrium
density $\rho_0$ by the condition \cite{Ruban2000}
\begin{equation}\label{divD}
(\nabla_{\bf R}\cdot{\bf D}({\bf R}))=\rho_0(\g{R}).
\end{equation}

{\bf 5.}
Unfortunately, practical use of the new variables $\g{R}(\nu,\xi,t)$ is
rather difficult in general case because of necessity to find the potential
component of the canonical momentum field $\g{p}$ from the system of
equations (\ref{Omega_def}) and (\ref{density}) for substitution of
$\g{p}\{\g{\Omega}\}$ into the equation (\ref{Hdef}). However, in analysis
of some situations this difficulty can be avoided. For instance, the vortex
line representation allows in the most simple way to study the so called
local induction approximation (LIA) in dynamics of thin vortex filaments.
Let's note that in usual Eulerian homogeneous hydrodynamics the LIA gives
the integrable equation which is gauge equivalent to 1D nonlinear
Schroedinger equation \cite{Hasimoto}. Our purpose now is to derive the
local induction equation for a slender vortex filament in a relativistic
fluid. Let's suppose that the vorticity is present in the system in form of
a quasi-one-dimensional structure with the circulation 
$\Gamma=\int_{\cal N} d^2\nu$, with a typical width $d$ and with a typical
longitudinal scale $L\gg d$. Then we may neglect the dependence of the
vortex line shape on the label $\nu$ in the main approximation and consider
the vortex filament as the single curve $\g{R}(\xi,t)$. In such conditions,
the Hamiltonian ${\cal H}_*$ is determined by the close neighborhood of
this curve, so that with the logarithmic accuracy ${\cal H}_*$ is equal to
the following expression
\begin{equation}\label{LIA_ham}
{\cal H}_*\approx{\cal H}_\Lambda=\Gamma \Lambda
\oint\frac{\rho_0(R)A(R)}{B(R)}|\g{R}_\xi|d\xi, 
\end{equation}
where the constant $\Lambda$ contains the large logarithm $\ln(L/d)\gg 1$
\begin{equation}
\Lambda=\frac{\Gamma}{4\pi\lambda}\ln\left(\frac L d\right).
\end{equation}
With help of the equations (\ref{main}) and (\ref{LIA_ham}) we can obtain
the equation of motion for a thin vortex filament in a relativistic fluid
placed into a central-symmetric static gravitational field:
$$
[\g{R}_\xi\times\g{R}_t]\left(\frac{\rho_0}{\Lambda}\right)=
\nabla\left(\frac{\rho_0 A}{B}\right)\cdot |\g{R}_\xi|
-\partial_\xi\left(\left(\frac{\rho_0 A}{B}\right)
\cdot\frac{\g{R}_\xi}{|\g{R}_\xi|}\right).
$$
Being solved with respect to the time derivative and rewritten in terms of
the unit tangent vector $\g{t}$, the unit binormal vector $\g{b}$ and the 
curvature $\kappa$ of the line, the above equation takes the form
\begin{equation}\label{LIA}
\g{R}_t\cdot\left(\frac{B(R)}{A(R)}\right)\cdot\frac{1}{\Lambda}=
[\nabla\ln\left(\frac{\rho_0(R) A(R)}{B(R)}\right)\times\g{t}]+\kappa\g{b},
\end{equation}
The condition of its applicability, besides $L\gg d$, is also the
inequality
\begin{equation}\label{conditionGamma}
\frac{A}{\lambda B}\cdot\frac{\Gamma}{d}\ll\sqrt{\frac{A}{B}},
\end{equation}
which means the smallness of the maximal velocity in the flow in comparison
with the speed of light.

Let us say finally few words about possible situation when the static
distribution of the matter in space is highly inhomogeneous. Let's suppose
that the equation of state $\varepsilon(n)$ results in formation of a dense
and massive core of relatively small size $r_*$ and extended easy shell
of a mass being much smaller than the mass of the core $M_*$. Then, as far
as we are interested in the dynamics of a thin vortex filament in the
shell, one can belive that the functions $A(r)$ and $B(r)$ in the equation
(\ref{LIA}) at $r>r_*$ are the same as in the empty space around the mass
$M_*$ \cite{LL2}:
\begin{equation}\label{A}
A(r)=\left(1-\frac{M_*}{2r}\right)^2\left(1+\frac{M_*}{2r}\right)^{-2},
\end{equation}
\begin{equation}\label{B}
B(r)=\left(1+\frac{M_*}{2r}\right)^4.
\end{equation}
The density $\rho_0(r)$ at $r>r_*$ in this case is determined by the
hydrostatic equation (\ref{hydrostat}) with the given $A$ and $B$.

\bigskip

This work was supported by RFBR (grant 00-01-00929), by
Program of Support of the Leading Scientific Schools (grant 00-15-96007), 
by the INTAS (grant 96-0413),
and by the Fund of Landau Postdoc Scholarship
(KFA, Forschungszentrum, Juelich, Germany).

\end{document}